# Ultra-broadband surface-normal coherent optical receiver with nanometallic polarizers

Go Soma[1,*], Warakorn Yanwachirakul[1], Toshiki Miyazaki[1], Eisaku Kato[1], Bunta Onodera[1], Ryota Tanomura[1], Taichiro Fukui[1], Shota Ishimura[2], Masakazu Sugiyama[3], Yoshiaki Nakano[1] & Takuo Tanemura[1,*]

[1] School of Engineering, The University of Tokyo, 7-3-1 Hongo, Bunkyo-ku, Tokyo 113-8656, Japan

[2] KDDI Research Inc., 2-1-15 Ohara, Fujimino-shi, Saitama 356-8502, Japan

[3] Research Center for Advanced Science and Technology, The University of Tokyo, 4-6-1 Komaba, Meguro-ku, Tokyo 153-8904, Japan

*Corresponding authors: tanemura@ee.t.u-tokyo.ac.jp and soma@hotaka.t.u-tokyo.ac.jp

**ABSTRACT**

A coherent receiver that can demodulate high-speed in-phase and quadrature signals of light is an essential component for optical communication, interconnects, imaging, and computing. Conventional waveguide-based coherent receivers, however, exhibit large footprints, difficulty in coupling a large number of spatial channels efficiently, and limited operating bandwidth imposed by the waveguide-based optical hybrid. Here, we present a surface-normal coherent receiver with nanometallic-grating-based polarizers integrated directly on top of photodetectors without the need for an optical hybrid circuit. Using a fabricated device with the active section occupying a 70-µm-square footprint, we demonstrate demodulation of high-speed (up to 64 Gbaud) coherent signals in various formats. Moreover, ultra-broadband operation from 1260 nm to 1630 nm is demonstrated, thanks to the wavelength-insensitive nanometallic polarizers. To our knowledge, this is the first demonstration of a surface-normal homodyne optical receiver, which can easily be scaled to a compact two-dimensional arrayed device to receive highly parallelized coherent signals.

High-speed coherent optical receivers that can detect the complete amplitude and phase information of light are becoming increasingly more important in optical communication and interconnects. With the rapid spread of cloud computing, video streaming, and Internet of Things (IoT) services, the global data traffic is raising exponentially with an annual growth rate of ~60% [1-5]. To accommodate this ever-growing bandwidth, there is a strong need to use highly parallelized superchannels in both wavelength and space dimensions [2,3]. Indeed, Peta-bit/s coherent optical transmission experiments have been demonstrated successfully by using more than 10,000 wavelength-division-multiplexed (WDM) and spatial-division-multiplexed (SDM) channels through multicore and multimode fibers [6,7]. Such parallelized systems also enable substantial savings in power consumption and receiver cost by performing the digital signal processing (DSP) functions jointly across the multiple coherent superchannels [8,9]. Moreover, utilization of parallelism naturally follows the scaling trend of the multicore microprocessors and can potentially realize ultralow-energy optical interconnects by eliminating the need for the timing and serialization electrical circuits [10]. To implement these technologies in actual systems, dense integration of massive optical transponders in a compact footprint is indispensable. As for the transmitter, ultra-compact and high-speed in-phase and quadrature (IQ) modulator arrays using plasmonic waveguides have been demonstrated [11,12]. With the length of only a few tens of micrometers, these modulator arrays can be connected directly to multicore fibers. As the counterpart, therefore, ultra-compact and broadband coherent receiver arrays are required, that can be integrated densely to receive highly parallelized optical signals.

In addition to optical communication and interconnects, densely integrated coherent receivers are required in diverse emerging applications, including three-dimensional (3D) imaging [13-15] and deep neural networks [16]. For example, a large-scale two-dimensional (2D) array of coherent detector pixels can retrieve 3D images at high accuracy and sensitivity, enabling compact and low-cost 3D imaging cameras used for robotics and autonomous navigations systems [13]. On the other hand, a massive number of high-speed coherent receivers can also be used as quantum photoelectric multipliers, which are scalable in both the temporal and spatial domains. This allows us to construct ultralow-energy highly scalable optical neural networks with a large number ($>10^6$) of neurons needed for practical deep-learning applications [16].

Conventionally, coherent optical receivers have been implemented by waveguide-based photonic integrated circuits (PICs) on various platforms, including silica [17], indium phosphide (InP) [18-20], and silicon (Si) [21-24]. Among them, the silicon photonic platform offers the highest optical confinement. Nevertheless, a single-channel single-polarization coherent receiver typically occupies a millimeter-square-scale footprint to contain a complicated PIC with an optical coupler (either a grating coupler or an edge coupler with a spot-size converter), a 90° optical hybrid, four photodetectors (PDs), and optical waveguides to connect them. In addition, the chromatic dispersion of the waveguide-based optical hybrid inherently makes the phase adjustment among the four output ports to be wavelength-sensitive, which fundamentally limits the operating bandwidth of the receiver to typically around 100 nm or less even with a careful design [23-25]. Finally, efficient simultaneous coupling of multiple optical signals from a multicore or multimode fiber to such waveguide-based arrayed device would become increasingly more challenging as the number of spatial channels increases [26].

In this paper, we propose and experimentally demonstrate a surface-normal homodyne coherent optical receiver, which can solve the scaling issues of the current coherent receivers. By integrating compact polarizers based on subwavelength nanometallic gratings directly onto a PD array, the required functionality of optical hybrid is realized without using a complicated PIC. As a result, the footprint of the entire device is determined simply by the detection area of four PDs, which can, in principle, be on a scale of tens of micrometers. The surface-normal configuration naturally allows the extension to a large-scale arrayed device to receive hundreds of spatial channels directly from a multicore fiber with minimal insertion loss and footprint. Dual-polarization detection of multiple-channel signals can also be realized by simply implementing the polarization-diversity architecture using a single polarization beam splitter (PBS), irrespective of number of channels. Furthermore, the phase property of the optical hybrid is determined simply by the geometrical angles of the nanometallic polarizers, independent of the wavelength. The proposed device would, therefore, be an ideal candidate for receiving highly parallelized ultra-broadband WDM-SDM signals in the future optical communication and interconnect systems.

To prove the concept, we fabricate an InGaAs-based pin PD array with integrated Au wire-grid polarizers. Using the fabricated device, we experimentally demonstrate high-speed coherent optical signal detection in various formats, such as 64 Gbaud quadrature phase-shift keying (QPSK), 60 Gbaud 8-ary quadrature amplitude modulation (8QAM), and 50 Gbaud 16QAM signals. Ultra-broadband operation covering all the wavelength ranges from 1280 nm to 1630 nm is then confirmed. While a heterodyne-based receiver array [27] and the passive optical hybrids without detectors [28,29] in surface-normal configurations have been reported previously, a full homodyne coherent receiver chip with integrated balanced detectors has not been realized experimentally. To the best of our knowledge, this is the first demonstration of a surface-normal homodyne coherent optical receiver.

**Results**

**Surface-normal coherent receiver concept.** The proposed surface-normal coherent receiver is illustrated in Fig. 1(a). It consists of four PDs attached with subwavelength nanometallic gratings oriented in different angles. When designed properly, each grating functions as a wire-grid polarizer that transmits only the transverse-magnetic (TM) component, which has the electric field aligned orthogonal to the wires [30-32]. The signal and the local-oscillator (LO) lightwaves are polarized in the left- and right-handed circular states, respectively, and are incident to the device at a normal angle. After transmitting through the wire-grid polarizer, they interfere with an optical phase difference represented as two times the polarizer angle (see Supporting Information, Note 1 for the detailed theoretical derivation). Therefore, a 90° optical hybrid can be realized by setting the angles of four polarizers to be 0°, 45°, 90°, and 135°. Then, from the differential photocurrent signal between the paired PDs with 0° and 90° (45° and 135°) polarizers, we can detect the real (imaginary) part of the signal. We should note that a similar concept of retrieving a full complex optical field using angled polarizers has been recognized for digital holography [33,34]. Here, we apply this scheme to the high-speed surface-normal coherent receiver for the first time.

We employ Au subwavelength grating with a thin Ti adhesion layer as a compact wire-grid polarizer as shown in Fig. 1(a) inset, which also serves as a low-electrical-resistance top contact of the InGaAs-based pin PD. The device was fabricated by defining the Au grating structures using the electron-beam lithography and reactive-ion-etching processes (see Methods for the details of device fabrication). Figures 1(b)-(d) show the microphotograph and the scanning electron microscope (SEM) images of the fabricated device.

Unlike the conventional waveguide-based coherent receiver, where the optical hybrid is implemented by a waveguide-based coupler or interferometer, the proposed scheme is inherently broadband since the optical phase difference between the signal and LO is determined precisely by the geometrical angle of the metallic grating attached on each PD. In addition, the wire-grid polarizer generally operates in ultrawide wavelength ranges [32]. Furthermore, the use of metallic gratings functions as low-electrical-resistance electrodes, in which case the electrical bandwidth is determined merely by the parasitic capacitance of the PD. Although we employ standard pin PD in this work for proof-of-concept demonstration, sub-wavelength stripe-antenna-based PDs based on hot electrons [35,36] or metal-semiconductor-metal (MSM) photodetectors [37] with plasmonic resonances can also be used to shrink the PDs to nanometer scale to realize ultra-low-capacitance high-speed receivers.

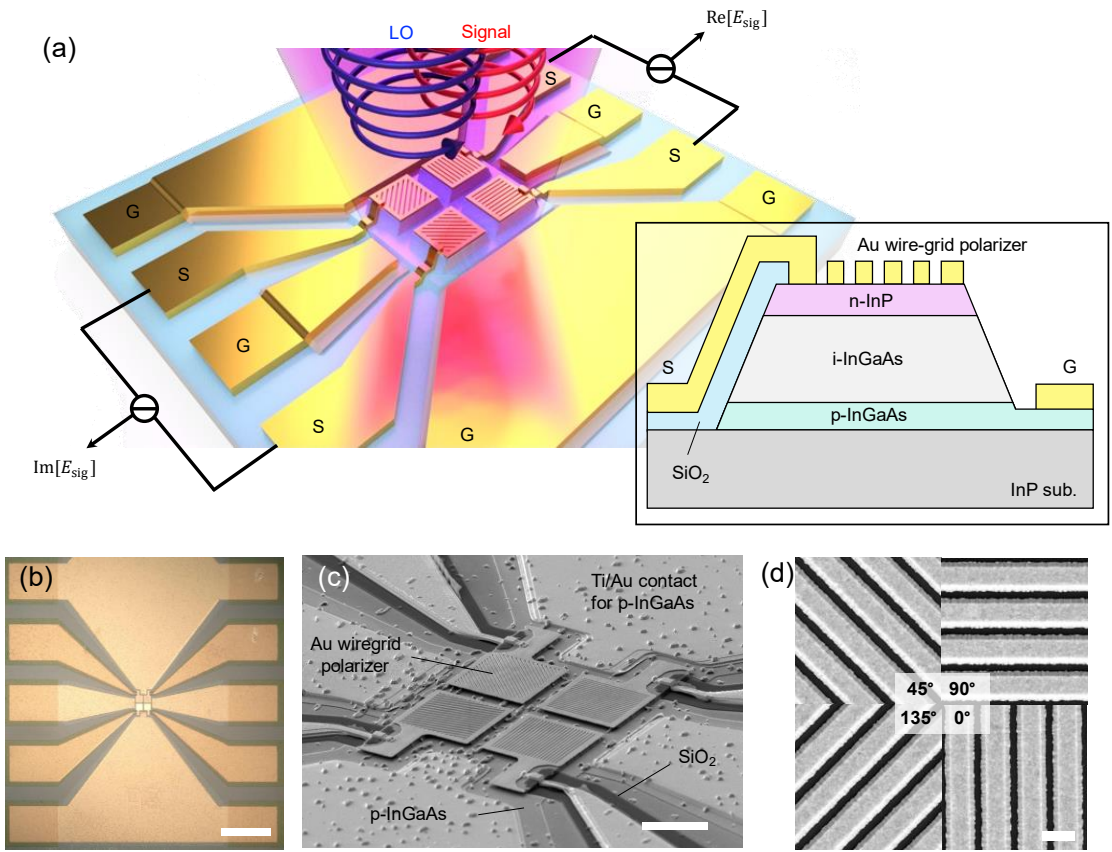

**Figure 1.** Proposed surface-normal coherent receiver with nanometallic-polarizer-integrated PDs. (a) Schematic of the device. The inset shows the cross-section of the actual fabricated device. The device is composed of four PDs with integrated polarizers, which have different angles of 0°, 45°, 90°, and 135°. From the differential photocurrent signal between the paired PDs with 0° and 90° (45° and 135°) polarizers, the real (imaginary) part of the signal is obtained. The polarizer is realized by integrating a subwavelength Au grating directly on top of InGaAs-based pin PD. (b)-(d) Microphotograph (b) and SEM images (c,d) of the fabricated device. Scale bars in (b), (c), and (d) are 200 µm, 20 µm, and 1 µm, respectively.

**Design and characterization of Au wire-grid polarizer.** The wire-grid polarizer used in this work consists of subwavelength Au grating as shown in Fig. 2(a). Since the gap is narrower than half the wavelength, only an evanescent wave can be excited when TE light with the electric field oriented along the wires is incident. As a result, the transmittance of TE light, $T_{TE}$, reduces exponentially as the Au thickness $t$ increases (Fig. 2(b)). On the other hand, when TM light with the electric field aligned perpendicular to the wires is input, a metal-insulator-metal (MIM) mode is excited that propagates through the gap and shows a Fabry-Perot-like behavior due to the reflections at the two interfaces. Thus, the transmittance of the TM mode, $T_{TM}$, oscillates periodically with $t$ as shown in Fig. 2(b). By adjusting the Au thickness properly, therefore, this structure operates as an ideal polarizer that transmits only the TM component of the incident light.

To derive the optimal design of the Au wire-grid polarizer depicted in Fig. 2(a), the transmittance properties are simulated for the TE and TM input cases by the 2D finite-difference time-domain (FDTD) method (Ansys Lumerical FDTD). Then, the thickness ($t$), width ($w$), and pitch ($a$) of the grating are tuned to maximize the difference in transmittance between the TE and TM modes. As a result, we derive the optimal parameters as $t$ = 260 nm, $w$ = 700 nm, and $a$ = 1100 nm (see

Supporting Information, Note 2 for details). Figure 2(c) shows the simulated electric field distribution Re[$E_y$] and Re[$E_x$] for the TE and TM inputs at a wavelength of 1550 nm.

Figure 2(d) shows the experimentally measured responsivities of the fabricated PD with the integrated wire-grid polarizer. Ultra-wideband characteristics with a flat responsivity for the TM input and large polarization extinction ratio are obtained from 1300 nm to 1600 nm. This result implies that our surface-normal coherent receiver should operate over an ultrabroad wavelength range, covering from O-band to L-band.

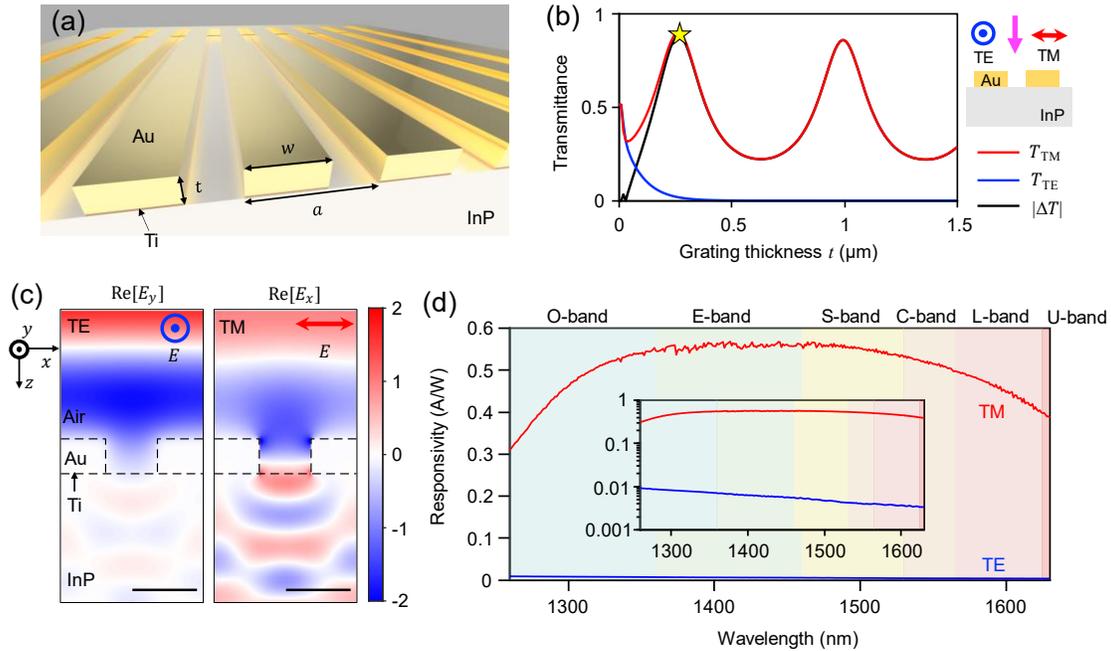

**Figure 2.** Design and measurement of the Au wire-grid polarizer. (a) Subwavelength Au grating with a 10-nm Ti adhesion layer. (b) Simulated transmittance for TE ($T_{TE}$) and TM ($T_{TM}$) modes and their difference ($\Delta T$) as a function of the Au thickness (wavelength = 1550 nm, $a$ = 1100 nm, $w$ = 700 nm). The star represents the optimal thickness used in the fabricated device. (c) Simulated electric field distribution Re[$E_y$] and Re[$E_x$] for the TE and TM inputs for the optimized structure

(wavelength = 1550 nm, $a$ = 1100 nm, $w$ = 700 nm, $t$ = 260 nm). For clarity, the large field at the edges of Au is saturated in these color plots. The scale bar is 500 nm. (d) Measured responsivity of the fabricated PD with the integrated polarizer for TE and TM inputs. Inset is the log-scale plot of the same experimental data.

**High-speed coherent detection experiment.** Using the fabricated receiver shown in Fig. 1(b), the coherent signal detection experiment was carried out at a 1550-nm wavelength. The setup is shown in Figs. 3(a) and (b) (see Method for details). Figure 3(c) shows the measured bit-error-rates (BERs) of 12.5 Gbaud QPSK, 8QAM, and 16QAM signals plotted as a function of the optical signal-to-noise ratio (OSNR). The results are comparable to those obtained using a commercial waveguide-based coherent receiver (Fraunhofer HHI: CRF-25) (see Supporting Information, Note 3 for details). In all cases, OSNR-limited behavior with the BER below $10^{-4}$ is confirmed. The measured constellation diagrams for respective cases are shown in Figs. 3(d)-(f).

Figure 3(g) shows the measured BERs as we increase the baudrate up to 64 Gbaud for different modulation formats. The received OSNR is fixed to 21.7 dB in this experiment. The constellation diagrams for various cases are shown in Figs. 3(h)-(m). Although the BER increases with the baudrate due to the reduced OSNR margin, the BERs below the soft-decision forward-error-correction (SD-FEC) limit of $4\times10^{-2}$ are obtained in all cases: 64 Gbaud QPSK, 60 Gbaud 8QAM, and 50 Gbaud 16QAM.

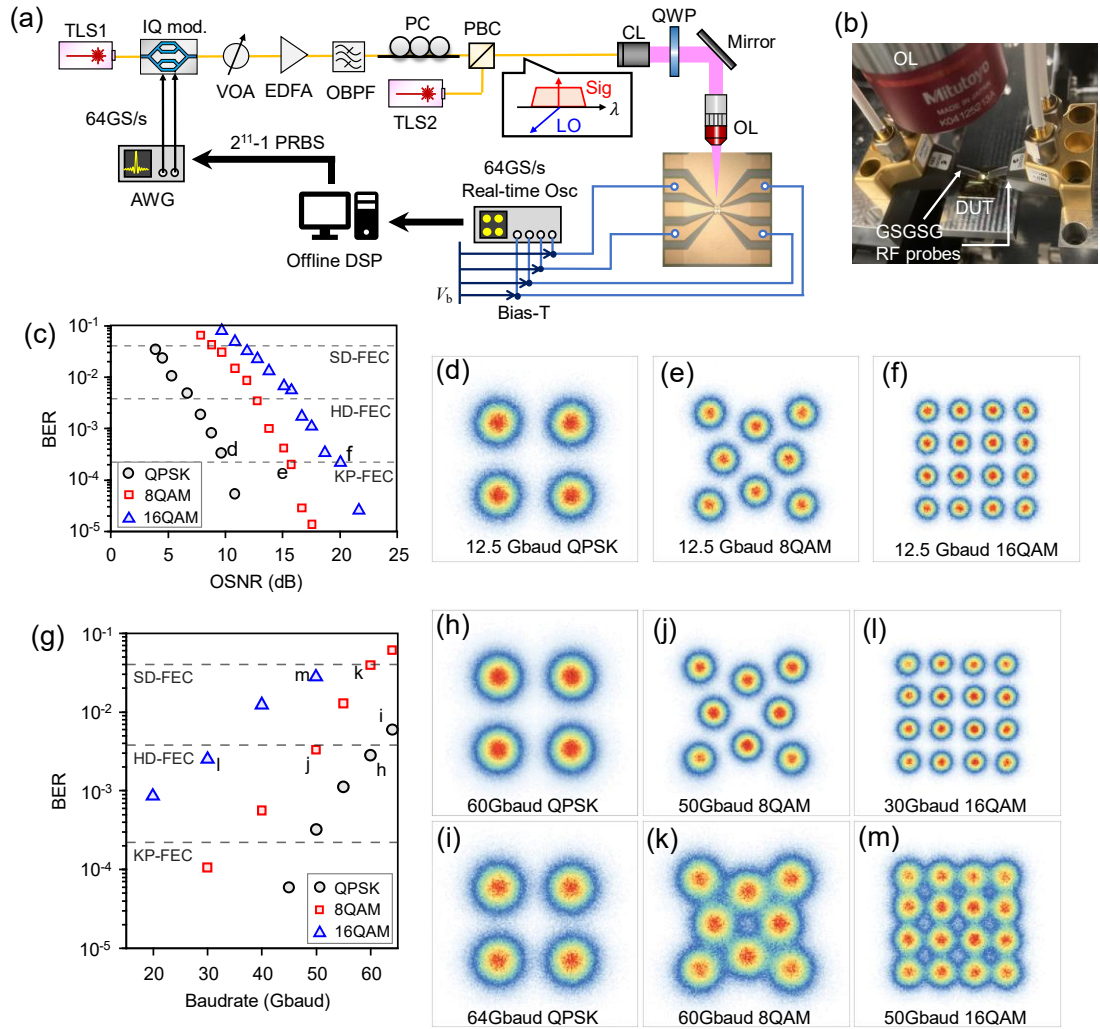

**Figure 3.** Coherent detection experiment at 1550 nm wavelength. (a) Experimental setup. TLS: tunable laser source, IQ mod: IQ modulator, AWG: arbitrary waveform generator, VOA: variable optical attenuator, EDFA: erbium-doped fiber amplifier, OBPF: optical bandpass filter, PC: polarization controller, PBC: polarization beam combiner, CL: collimation lens, QWP: quarter-wave plate, OL: objective lens, DUT: device under test, Osc: oscilloscope, DSP: digital signal processing. (b) Photograph of the setup. Four electrical signals are collected by two GSGSG RF probes. (c) Measured BERs for 12.5 Gbaud QPSK, 8QAM, and 16QAM signals as a function of the received OSNR. (d)-(f) Measured constellation diagrams of 12.5 Gbaud (d) QPSK, (e) 8QAM

and (f) 16QAM signals. (g) Measured BERs with increasing baudrate for QPSK, 8QAM, 16QAM signals. The received OSNR is fixed to 21.7 dB in all cases. (h)-(m) Measured constellation diagrams of (h) 60 Gbaud QPSK, (i) 64 Gbaud QPSK, (j) 50 Gbaud 8QAM, (k) 60 Gbaud 8QAM, (l) 30 Gbaud 16QAM, and (m) 50 Gbaud 16QAM signals.

**Ultrabroad wavelength range experiment.** As discussed above, the proposed surface-normal coherent receiver is inherently broadband owing to the wideband characteristic of the wire-grid polarizer (shown in Fig. 2(d)) as well as the intrinsic wavelength insensitivity of the optical hybrid, where the relative optical phase is defined by the geometrical angles of the polarizers. To confirm this property, the performance of the receiver is compared at different wavelength bands. The experimental setup is shown in Fig. 4(a) (see Method for details). Due to the lack of multiple tunable laser sources (TLSs) in all the wavelength bands, the self-coherent scheme is employed in this measurement; the LO laser source is shared with the signal source.

Figure 4(b) shows the measured BERs of 12.5 Gbaud QPSK signals at 1310 nm, 1480 nm, and 1550 nm as a function of the received optical power. In all wavelengths, we obtain BERs well below the FEC limits. The difference in the received power sensitivity is attributed to the LO power variation depending on the wavelength. Figure 4(c) shows the obtained BERs in the entire wavelength range from 1280 nm to 1630 nm. The power of the signal and LO beat signal $\sqrt{P_{sig}P_{LO}}$ is also plotted. We can see that the BER degrades in some wavelength ranges due to the insufficient $\sqrt{P_{sig}P_{LO}}$ available from the TLS. Nevertheless, low BERs below the SD-FEC limit of $4\times10^{-2}$ are confirmed over the entire wavelength range from 1280 nm to 1630 nm, which was only limited by the operating wavelength range of the available laser sources.

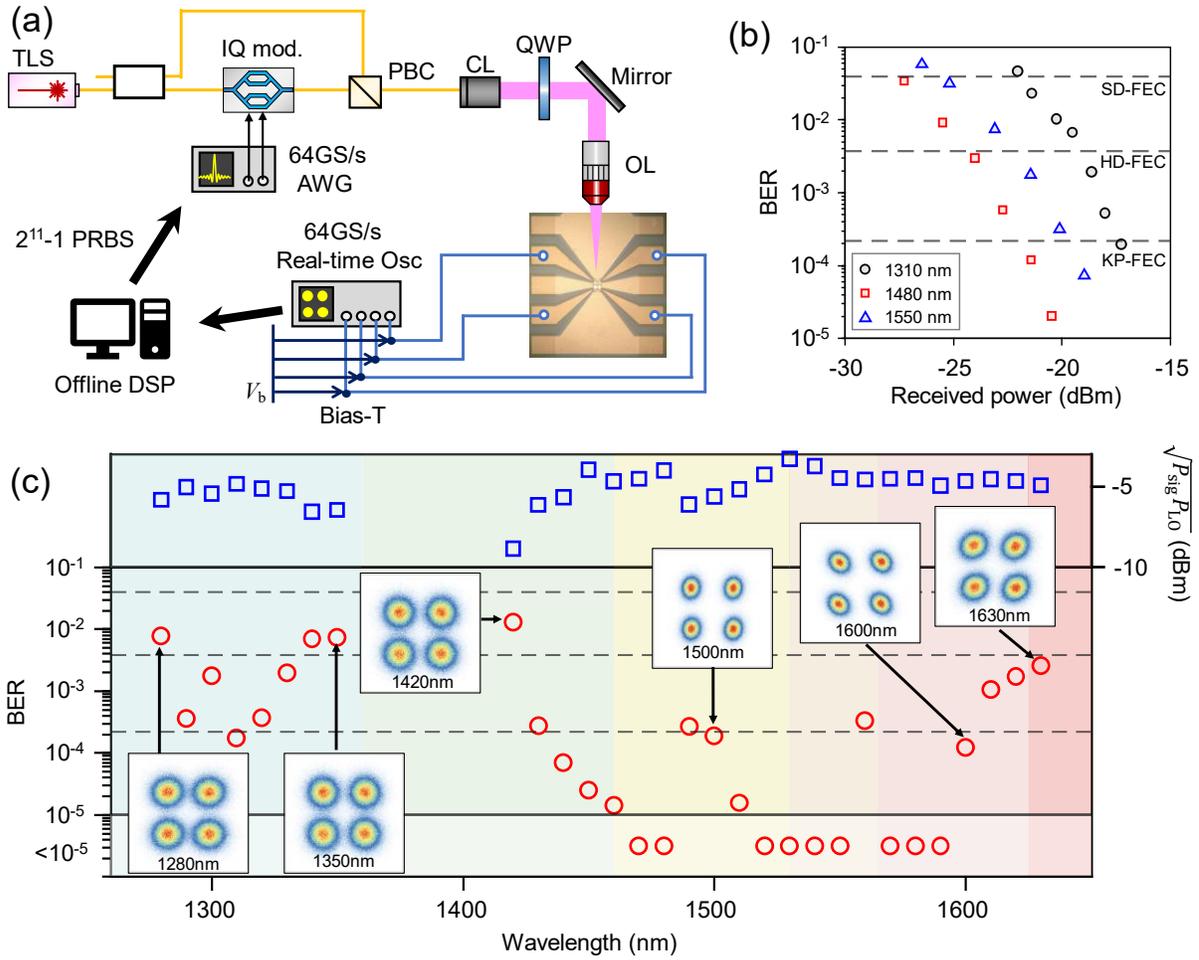

**Figure 4.** Coherent detection experiment at different wavelength bands. (a) Experimental setup. (b) Measured BER characteristics of 12.5 Gbaud QPSK signal at 1310 nm, 1480 nm, and 1550 nm. (c) Measured BERs of 12.5 Gbaud QPSK signal (red circles) and $\sqrt{P_{sig}P_{LO}}$ (blue squares) at different wavelengths. The constellation diagrams at several representative wavelengths are shown in the inset.

**Conclusion**

In summary, we have proposed and demonstrated a surface-normal homodyne coherent optical receiver for the first time. With the nanometallic-grating-based polarizers integrated directly on top of a high-speed PD array, balanced homodyne detection was performed without a need for a complicated optical hybrid circuit. Using a fabricated device with an active section occupying only a 70-µm-square footprint, demodulation of 64 Gbaud QPSK, 60Gbaud 8QAM, and 50Gbaud 16QAM signals was demonstrated at 1550-nm wavelength. In addition, since the phase property of the optical hybrid could be controlled simply by the geometrical angles of the nanometallic gratings, ultra-broadband operation from 1280 nm to 1630 nm was achieved experimentally.

The surface-normal configuration of the demonstrated coherent receiver enables natural extension to a large-scale 2D array to receive highly parallelized coherent signals in a minimal footprint. While we employed a conventional InGaAs-based pin PD structure for detection in this work, a more compact and lower-capacitance structure could also be used [35-37] to further enhance the spatial density and the electrical bandwidth. Moreover, our device concept is applicable to other material platforms as well, such as germanium-based PDs integrated on silicon to realize a complementary metal-oxide-semiconductor (CMOS) compatible low-cost surface-normal coherent receiver array.

While we have only demonstrated single-polarization detection in this work, a dual-polarization coherent receiver can readily be implemented by using two sets of the devices and a PBS to construct a polarization-diversity architecture. We should note that only a single PBS is required, irrespective of the number of parallelized signals. The additional complexity to achieve dual-polarization operation would, therefore, become nominal if we integrate hundreds of channels on a compact chip. The demonstrated surface-normal coherent receiver would provide a new paradigm in diverse fields of photonics, including optical communication, interconnects, imaging, and computing.

**Methods**

**Device fabrication**. The epitaxial layers, consisting of p-InGaAs (250 nm), i-InGaAs (800 nm), and n-InP layers (250 nm), were grown on a semi-insulating InP substrate by the metal-organic chemical vapor deposition (MOCVD). The Ti/Au metallization layers and the 200-nm-thick $SiO_2$ hard mask layer were deposited by radio-frequency (RF) sputtering. The grating patterns were defined by an electron-beam writing system with ZEP520A resist, which were first transferred to the $SiO_2$ layer by inductively-coupled-plasma reactive-ion etching (ICP-RIE) using $CHF_3$. Then, the Au and Ti layers were etched by ICP-RIE using Ar. After removing the $SiO_2$ hard mask by buffered hydrofluoric acid (BHF), the PD mesas were formed by the wet chemical etching with $HCl + H_3PO_4$ for InP layers and $H_2SO_4 + H_2O_2$ for InGaAs layers. Then, the $SiO_2$ insulation layer was formed by plasma-enhanced chemical vapor deposition (PECVD) and wet-etched by BHF after photolithography for contact opening. Finally, the Ti/Au electrodes were formed by the liftoff process. Two types of devices with the active PD areas of 200-μm-square and 30-μm-square were fabricated on the same chip for the static and high-speed experiments, respectively.

**High-speed coherent detection experiment.** The setup for the high-speed experiment is shown in Fig. 3(a). In this experiment, we used a device with four 30-μm-square PDs (70-μm-square total footprint). A CW light at a 1550-nm wavelength from a TLS was modulated by a $LiNbO_3$ IQ modulator, which was driven by an arbitrary waveform generator (AWG) running at 64 GS/s. The Nyquist filter was applied to the driving electrical signals. The optical signal-to-noise ratio (OSNR) of the received signal was controlled by a variable optical attenuator (VOA) followed by an erbium-doped fiber amplifier (EDFA). The ASE noise outside the signal bandwidth was eliminated by an optical bandpass filter (OBPF). The LO from another TLS was combined with the signal by a polarization beam combiner (PBC). They were then converted to orthogonal circular polarization states via a quarter-wave plate (QWP) and then incident to the device. The four photocurrent signals from the PDs were collected through two GSGSG probes, as shown in Fig. 2(b). The bias voltage of -7 V was applied to all PDs through bias-tees. The four electrical signals were then captured by a real-time oscilloscope running at 128 GS/s for receiving signals above 60 Gbaud and at 64 GS/s in other cases. The captured signals were converted into in-phase and quadrature signals by taking the difference and demodulated through offline DSP. The signals were resampled and equalized with half-symbol-spaced adaptive finite-impulse-response (FIR) filters. As the adaptive algorithm, we employed the decision-driven least-mean-square (DD-LMS) algorithm, which was modified from that reported in [38,39], to remove the phase noise, the frequency offset, and the residual IQ imbalance.

**Ultrabroad wavelength range experiment.** The experimental setup is shown in Fig. 4(a). Due to the lack of two TLSs in some wavelength ranges, the light from a single TLS was split for generating the signal and the LO and combined by a PBC to construct a self-homodyne detection system. The roll-off factor of the raised cosine filter was set to 1. The received signal power was varied by adjusting the driving voltage from the AWG. We employed different sets of TLS, 3-dB coupler, and PBC for the three wavelength ranges of 1280-1350 nm, 1420-1500 nm, and 1510-1630 nm. The rest of the setup was the same as Fig 3(a).

## ASSOCIATED CONTENT

**Supporting Information**

The Supporting Information is available free of charge.

Details of theoretical derivation, optimization of wire-grid polarizer, and experimental comparison with a commercial coherent receiver.

## AUTHOR INFORMATION

**Corresponding Authors**

Takuo Tanemura – The University of Tokyo, 7-3-1 Hongo, Bunkyo-ku, Tokyo 113-8656, Japan

E-mail: tanemura@ee.t.u-tokyo.ac.jp

Go Soma – The University of Tokyo, 7-3-1 Hongo, Bunkyo-ku, Tokyo 113-8656, Japan

E-mail: soma@hotaka.t.u-tokyo.ac.jp

**Author Contributions**

G.S. conceived the concept, designed and developed the devices, designed and performed the experiments, and analyzed the data. W.Y. performed the MOCVD crystal growth. T.M., B.O., R.T., and E.K. contributed to the device fabrication. T.M. and T.F. contributed to building the measurement setup. S.I. contributed to the coherent detection experiment and the data analysis. M.S. and Y.N. contributed to developing the fabrication and experimental infrastructures. T.T. supervised and coordinated the project. G.S and T.T. wrote the manuscript. All authors discussed the results and commented on the manuscript.

**Funding Source**

National Institute of Information and Communications Technology (NICT), Japan.

ACKNOWLEDGMENT

This work contains the results obtained from a project commissioned by National Institute of Information and Communications Technology (NICT), Japan. A part of the device fabrication was conducted at the cleanroom facilities of d.lab in the University of Tokyo, supported by MEXT Nanotechnology Platform, Japan. The authors thank M. Fujiwara, A. Mizushima, and Y. Miyatake for the fabrication support. G.S. acknowledges the financial support from Optics and Advanced Laser Science by Innovative Funds for Students (OASIS).

*Supporting Information for*

# Ultra-broadband surface-normal coherent optical receiver with nanometallic polarizers

Go Soma[1,*], Warakorn Yanwachirakul[1], Toshiki Miyazaki[1], Eisaku Kato[1], Bunta Onodera[1],

Ryota Tanomura[1], Taichiro Fukui[1], Shota Ishimura[2], Masakazu Sugiyama[3], Yoshiaki Nakano[1] &

Takuo Tanemura[1,*]

[1] School of Engineering, The University of Tokyo, 7-3-1 Hongo, Bunkyo-ku, Tokyo 113-8656, Japan

[2] KDDI Research Inc., 2-1-15 Ohara, Fujimino-shi, Saitama 356-8502, Japan

[3] Research Center for Advanced Science and Technology, The University of Tokyo, 4-6-1 Komaba, Meguro-ku, Tokyo 153-8904, Japan

*Corresponding authors: tanemura@ee.t.u-tokyo.ac.jp and soma@hotaka.t.u-tokyo.ac.jp

**Note 1: Derivation of transmission through a polarizer**

The Jones matrix of a polarizer oriented in an angle $\theta$ is described as

$$\mathbf{T}_\theta = \begin{pmatrix} \cos\theta & -\sin\theta \\ \sin\theta & \cos\theta \end{pmatrix} \begin{pmatrix} 1 & 0 \\ 0 & 0 \end{pmatrix} \begin{pmatrix} \cos\theta & \sin\theta \\ -\sin\theta & \cos\theta \end{pmatrix}$$

$$= \begin{pmatrix} \cos^2\theta & \sin\theta\cos\theta \\ \sin\theta\cos\theta & \sin^2\theta \end{pmatrix}. \tag{1}$$

When the signal with a right-handed circular polarization state $\mathbf{E}_R = \frac{1}{\sqrt{2}}(1, i)^\mathrm{T}$ is incident to this polarizer, its Jones vector after transmission is written as

$$\mathbf{E}_{\text{sig}} = \mathbf{T}_\theta \mathbf{E}_R = \frac{e^{i\theta}}{\sqrt{2}} \begin{pmatrix} \cos\theta \\ \sin\theta \end{pmatrix}. \tag{2}$$

Similarly, for the left-handed incident LO wave, we obtain

$$\mathbf{E}_{\text{LO}} = \frac{e^{-i\theta}}{\sqrt{2}} \begin{pmatrix} \cos\theta \\ \sin\theta \end{pmatrix} \tag{3}$$

after the polarizer. From Eqs. (2) and (3), we can confirm that both the signal and LO have an identical linear polarization state at angle $\theta$ with the optical phase shifts of $\pm\theta$, respectively. We can, therefore, provide a relative phase difference of $2\theta$ by adjusting the polarizer angle $\theta$.

**Note 2: Optimization of wire-grid polarizer**

Here, we optimize the wire-grid polarizer structure to maximize the difference of the transmittance between the TE and TM modes. Fig. S1 shows the transmittance of TE and TM modes and their difference simulated at a wavelength of 1550 nm. For each case of the grating width ($w$) and the pitch ($a$), the thickness of the Au layer ($t$) is adjusted to the first transmission peak of the FP-like resonance for the TM mode, as explained in Fig. 2b of the main manuscript. The transmittance of the TE mode (Fig. S1(a)) reduces as the grating gap ($= a - w$) decreases since the decay rate of the evanescent mode increases monotonically as the gap becomes narrower. In contrast, the transmittance of the TM mode (Fig. S1(b)) at the FP-like resonance peak has a small dependence on $a$ and $w$. This is because the reflectance of the MIM mode at the grating-air and grating-InP interfaces is dependent on the impedance matching conditions, which varies with the grating structure. In addition, the plasmonic loss becomes large when the localized plasmonic resonance at the InP-metal interface becomes strong. To achieve a large transmission difference $\Delta T$, therefore, it is essential to have the grating gap narrow enough to keep the transmittance for the TE mode to be low, while matching the reflectance at the two interfaces and minimizing the plasmonic loss for the TM mode. On the other hand, a larger grating period is preferred from the fabrication point of view. From Fig. S1(c) and the above considerations, we select the grating parameters as $a = 1100$ nm, $w = 700$ nm, $t = 260$ nm in this work.

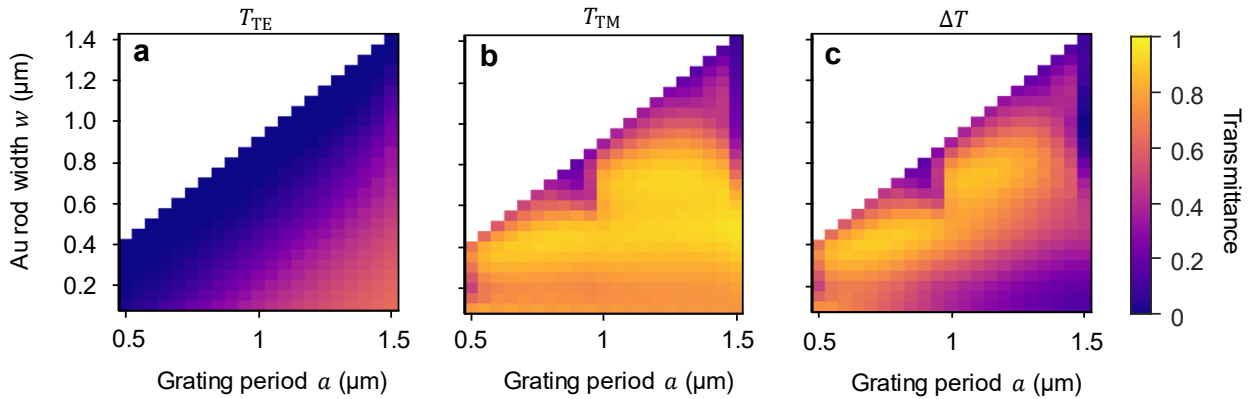

**Figure S1.** Simulated transmittance of TE (a) and TM (b) modes and their difference (c) at the wavelength of 1550 nm as a function of the grating width ($w$) and pitch ($a$). In each case, the thickness of the Au layer ($t$) is adjusted to the first FP resonance for the TM mode.

**Note 3: Comparison with commercial coherent receiver**

The performance of the fabricated device is compared with a commercial waveguide-based coherent receiver (Fraunhofer HHI, CRF-25). Figure S2 shows the measured BER characteristics of 12.5 Gbaud QPSK, 8QAM, and 16QAM signals detected by two receivers. In all cases, there is no significant difference between the two, indicating that the presented coherent receiver operates at the OSNR limit without a notable penalty.

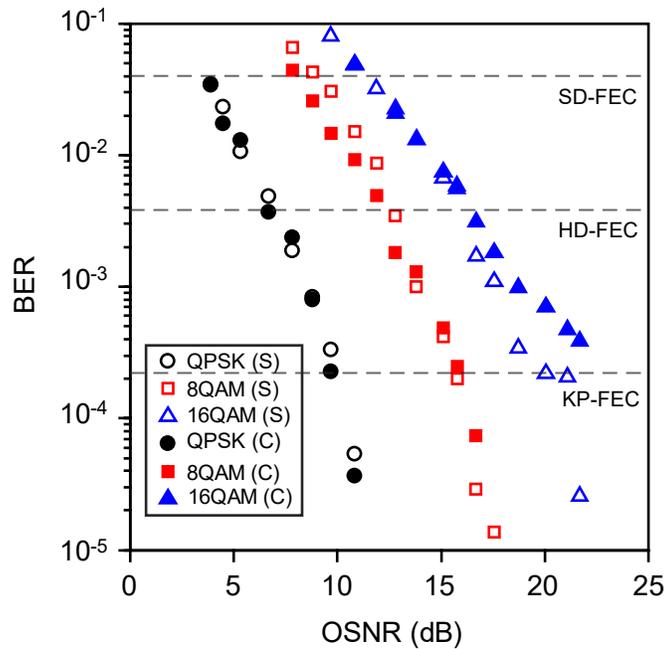

**Figure S2.** Comparison of BER characteristics for our surface-normal coherent receiver (S) and a commercial waveguide-based coherent receiver (C).